%% file: main.tex
\documentclass[nologo,a4paper,10pt]{ETHpaper}
\usepackage[british]{babel}
\usepackage{blkarray} \usepackage[square,numbers]{natbib}
\usepackage{bbold}

\title{Adapting to Disruptions: Flexibility as a Pillar of Supply Chain Resilience}
\author{Ambra Amico, Luca Verginer, Giona Casiraghi, Giacomo Vaccario, Frank Schweitzer}
\address{Chair of Systems Design, \\ ETH Zurich, Weinbergstrasse 56/58, 8092 Zurich, Switzerland \\[2mm] }
\www{\url{http://www.sg.ethz.ch}}

\newcommand{\veryshortarrow}[1][5pt]{\mathrel{\hspace{-.2em}\hbox{\rule[\dimexpr\fontdimen22\textfont2-.2pt\relax]{#1}{.4pt}}\mkern-4mu\hbox{\usefont{U}{lasy}{m}{n}\symbol{41}}\hspace{-.2em}}}
\renewcommand{\to}{\veryshortarrow}

\begin{document}

\maketitle

\begin{abstract}
  Supply chain disruptions cause shortages of raw material and products.
  To increase resilience, i.e., the ability to cope with shocks, substituting goods in established supply chains can become an effective alternative to creating new distribution links.
  We demonstrate its impact on supply deficits through a detailed analysis of the US opioid distribution system.
  Reconstructing 40 billion empirical distribution paths,
  our data-driven model allows a unique inspection of policies that increase the substitution flexibility.
  Our approach enables policymakers to quantify the trade-off between increasing flexibility, i.e., reduced supply deficits, and increasing complexity of the supply chain, which could make it more expensive to operate.
  
\end{abstract}

\section*{Introduction}
\input{introduction.tex}

\input{results.tex}

\section*{Discussion}
\input{discussion.tex}

\bibliography{references,zoter_references}
\bibliographystyle{sg-bibstyle}

\pagebreak
\section{Material and Methods}
\input{methods.tex}

\end{document}

%% file: introduction.tex
The complexity of supply chains---connecting manufacturers, distributors, retailers, and final buyers---has increased over the past century, raising concerns about their resilience~\citep{burkholz2019-internationalcrop}.
Recent events such as the COVID-19 pandemic, the war in Ukraine, and US--China trade disputes have affected supply chains by severely disrupting the global \emph{distribution} of raw materials and goods.
Following the Covid-19 pandemic, the US administration declared the ``Public Health Supply Chain'' a top national security issue and is seeking ``new approaches to build diversity and \textit{flexibility}''~\citep{thewhitehouse2021-nationalstrategy}.
To do so, policymakers and firms must quantify and devise policies to improve resilience, which is the ability to mitigate shortages following sudden reductions in products' availability.

There are several ways to tackle product shortages~\citep{urahn2017-drugshortages,jonghthyrade2021-futureproofingpharmaceutical}.
However, only two responses are \emph{immediately} available: rationing and substitution.
While rationing may become necessary as the shortage deepens, substitution is typically the first choice as it impacts final buyers the least.
A distributor has two strategies to implement substitution:
(i) establishing relations with \emph{new} distributors, or (ii) leveraging \emph{existing} relations to obtain a substitute good.

The first strategy requires searching for new distributors and establishing new business relations, which may be costly and time-consuming~\cite{trent2005achieving}.
The second strategy requires relaxing product preferences by accepting substitute goods from existing upstream distributors.
Inspired by the seminal works of \citet{tang2008-powerflexibility,dolgui2018-rippleeffect,ivanov2014-rippleeffect}, we call this last strategy \emph{flexibility}.
We show that policies fostering flexibility can considerably alleviate shortages.

An ideal dataset to study the power of flexibility is ARCOS~\citep{ARCOS}.
It lists \emph{all} drug shipments from 2006 to 2014 in the \emph{US opioid distribution system} that has been often affected by shortages with dramatic consequences~\citep{tucker2020-drugshortage,woodcock2013-economictechnological,miller2021-vulnerabilitymedical,wang2019-crisisepidemic,hollingsworth2018-parenteralopioid}.
This dataset offers an unprecedented view of distribution at a systemic scale, which is unique in supply chain research~\citep{bier2020-methodsmitigating}.
With these data, we reconstruct 40 billion distribution paths connecting manufacturers to more than a thousand distributors and 200 000 final buyers, i.e., pharmacies, hospitals, and practitioners.
Based on the reconstructed paths, we develop and estimate a data-driven model to investigate (i) how supply shocks lead to shortages and (ii) how fostering flexibility mitigates them.

\begin{figure}
    \centering
    \includegraphics[width=.7\textwidth]{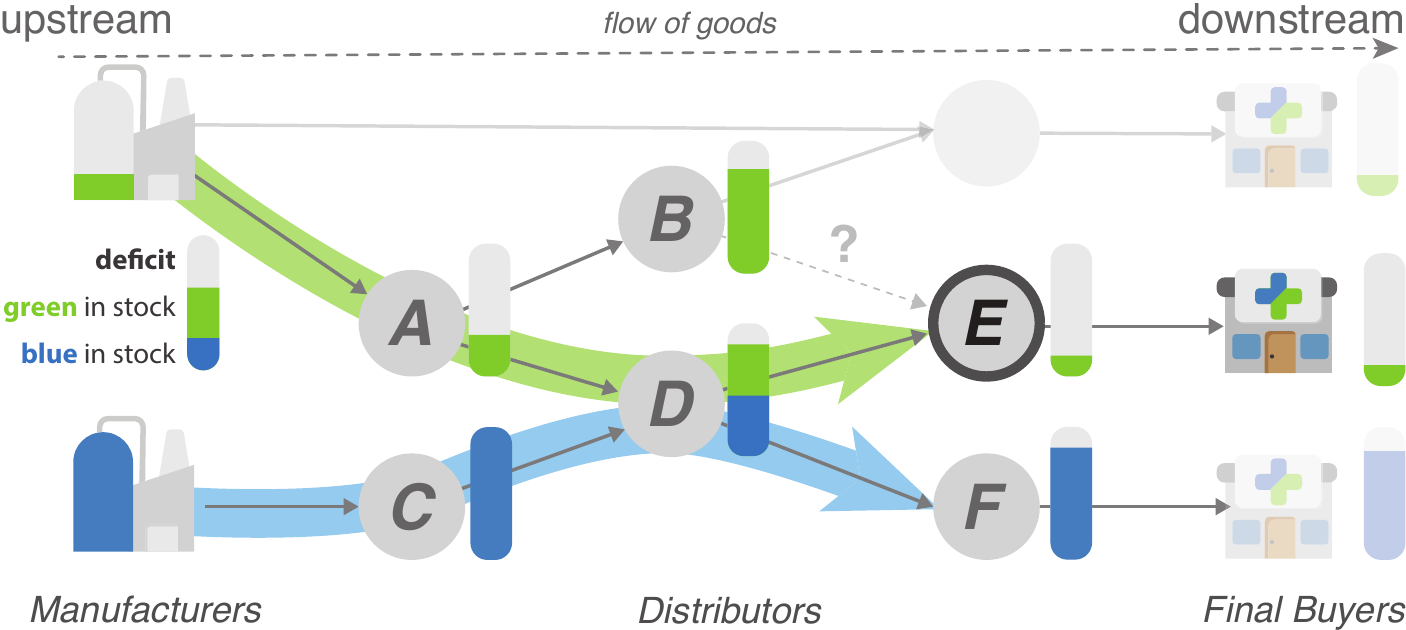}
    \caption{
        Schematic illustration of a distribution system of two perfectly substitutable products: blue and green.
        Goods flow from upstream to downstream distributors, and orders in the opposite direction.
        Grey arrows represent shipments of goods from one distributor to another.
        The two bold coloured arrows are \emph{distribution paths}, i.e., sequences of distributors through which goods arrive at their destination (final buyers).
        In this example, there is a shortage of green, shown by the deficit of green at both distributors and final buyers, while blue is fully available.
        Distributor $E$ has demand for green, exceeding the stock available upstream at $D$.
        $D$ could satisfy the demand with blue, a substitute.
        This is only possible if $E$ relaxes its upstream preference for green and accepts blue instead.
        Alternatively, $E$ could establish a new relation with $B$ to obtain green (dashed line).
        Assuming that the cost of establishing a new relation is higher than substituting blue for green, $E$ should choose the latter.
        This work focuses on this substitution, i.e., relaxing upstream preferences.
    }\label{fig:example}
\end{figure}

\section*{Upstream Preferences and Flexibility}
To operationalize flexibility in a distribution system, we focus on the \emph{distributor} of a good rather than the good itself.
To understand this change of perspective consider the example in \cref{fig:example}.
It shows  a distribution system of two substitutable goods: green and blue.
Distributor $E$ prefers goods coming from $A$ (green) over $C$ (blue).
We formalize these \emph{upstream preferences} as stochastic chains with memory~\citep{rissanen1983,buhlmann1999}.
These correspond to the probabilities that $E$ places an order to $D$ for goods coming from $A$ or $C$.
In this case, $\Pr(E\to D\to A)=1$ and $\Pr(E\to D\to C)=0$, respectively.
However, if $E$ had no specific preferences regarding $A$ (green) or $C$ (blue), it would instead receive goods solely based on their availability in $D$.
This implies that $E$ would adapt to the preferences of its upstream distributor $D$.
Thus, $\Pr(E\to D\to A) = \Pr(D\to A) = 0.5$ and $\Pr(E\to D\to C) = \Pr(D\to C) = 0.5$.
Flexibility $\phi_E$ is the propensity of distributor $E$ to relax its preferences in favour to those upstream.
Formally,
\begin{equation}\label{eq:example}
    \begin{aligned}
        \Pr(E\to D\to A | \phi_E) & := \phi_E \Pr(E\to D\to A) + (1-\phi_E) \Pr(D\to A)    \\
        \Pr(E\to D\to C | \phi_E) & := \phi_E \Pr(E\to D\to C) + (1-\phi_E) \Pr(D\to C)\;.
    \end{aligned}
\end{equation}
When $\phi_E>0$, distributor $E$ becomes more flexible in its preferences and starts sourcing goods from $C$, thus opening up an alternative distribution path: $C \to D \to E$.
Through this new path, $E$ can fulfill its demand by substituting the good (green) it needs with the substitute (blue) coming from $C$.
For instance, suppose E has a deficit of 4 green units, and D has a total stock of 4 units (2 green and 2 blue).
If $\phi_E=0$, E would only be able to fulfill 2 units of its demand by receiving 2 green units.
Instead, if $E$ partially relaxes its upstream preferences ($\phi_E=0.5$), $E$ could further reduce its deficit by an additional unit.

%% file: results.tex
\begin{figure}[t]\centering
    \begin{subfigure}{.48\textwidth}\centering
        \begin{subfigure}{\textwidth}\centering
            \includegraphics[width=\textwidth]{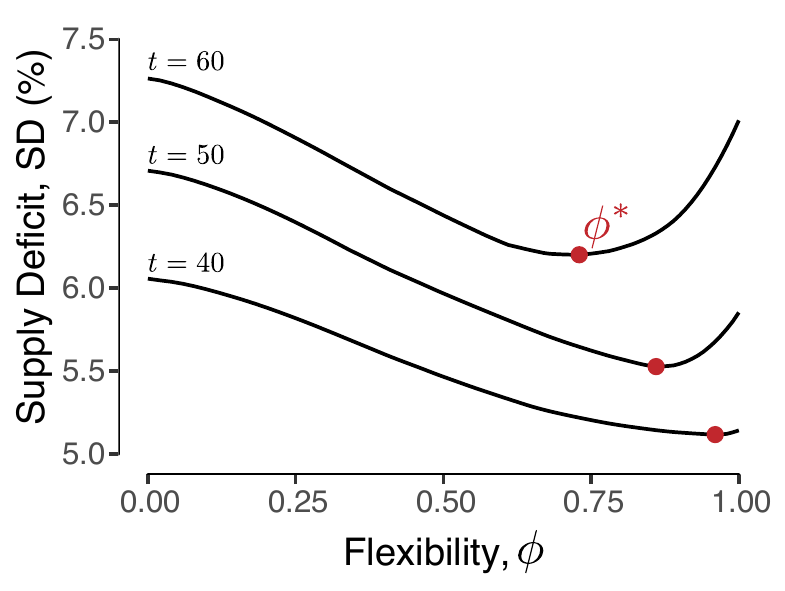}
            \caption{}\label{fig:deficit:phi_impact}
        \end{subfigure}\\
        \begin{subfigure}{\textwidth}\centering
            \includegraphics[width=\textwidth]{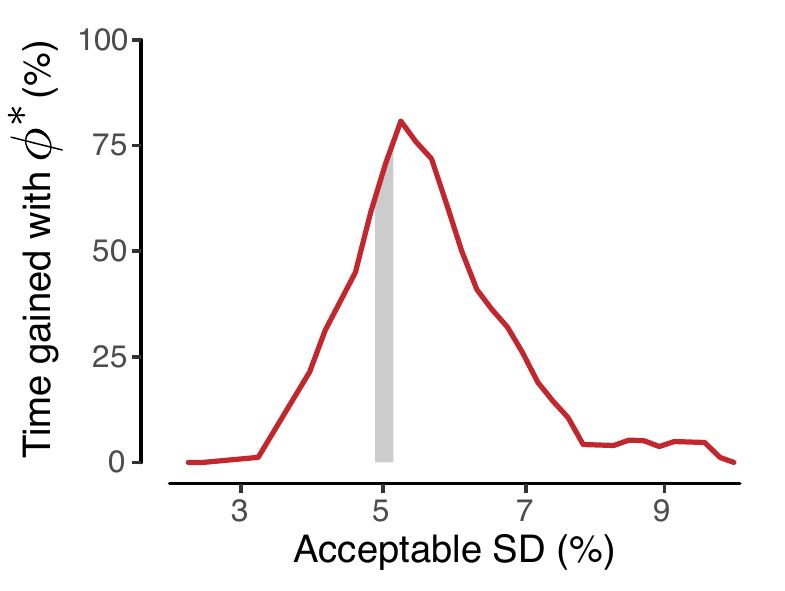}\addtocounter{subfigure}{1}
            \caption{}\label{fig:deficit:resupplygain}
        \end{subfigure}
    \end{subfigure}\hfill
    \begin{subfigure}{.48\textwidth}\centering
        \includegraphics[width=\textwidth]{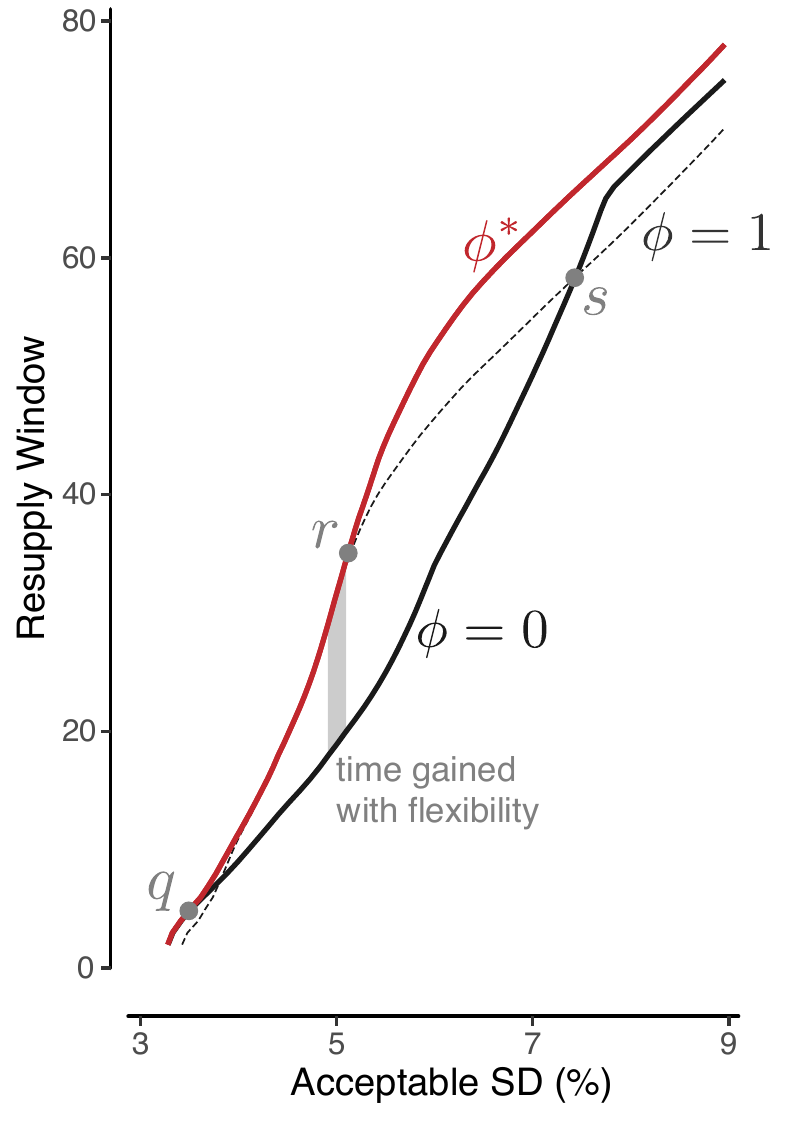}\addtocounter{subfigure}{-2}
        \caption{}\label{fig:deficit:resupplytime}
    \end{subfigure}
    \caption{
        Stress-test simulation results for the Oxycodone distribution system in 2012.
        \textbf{(a)} Percentage of final buyers' total demand that was not met, shown as the deficit, for flexibility values $\phi\in[0,1]$.
        The results are plotted at 40, 50, and 60 days after production stopped.
        The flexibility value $\phi^*$ yielding the maximum deficit reduction is shown in red.
        \textbf{(b)}
        Available resupply window shown as the time available to resupply before breaching an acceptable supply deficit (ASD), black line.
        The extended resupply window obtained with $\phi=\phi^*$ shown in red, is always above the black line.
        The resupply window obtained with full flexibility ($\phi=1$) is shown as the dashed line.
        Up to point $q$, the resupply window is the same for $\phi=0$ and $\phi^*$.
        Beyond point $r$, the largest resupply window is obtained for $\phi^*<1$.
        Beyond point $s$, full flexibility ($\phi=1$) is worse than no flexibility ($\phi=0$).
        \textbf{(c)} Time gained with flexibility, showing the increase in the time available for resupply for a given ASD.
    }\label{fig:deficit}
\end{figure}

\section*{Flexibility alleviates shortages}
\paragraph{Stress test of distribution systems}

We use stress test simulations to explore how the empirical distribution system may respond to supply shocks.
Using a data-driven agent-based model, we simulate distributors placing orders based on demand and distributing goods based on orders.
To simulate a sudden stop in production at $t=0$, we set upstream distributor stocks to only 70\% of their maximal capacity inferred from data.
We then analyze how this upstream deficit affects the final-buyers supply deficit and how the latter grows over time while production is halted.
See \cref{sec:methods:ARIO} for a detailed explanation of the simulations.

In \cref{fig:deficit}, we plot the outcome of the simulations for the Oxycodone distribution system.
\Cref{fig:deficit:phi_impact} shows final-buyers supply deficit at different times.
After 40 days, without flexibility, final buyers suffer a deficit of 6\%.
While this number might seem small, it corresponds to over $3$M missing Oxycodone doses.
This deficit continues to increase with time as the shortage remains unresolved and stocks are depleted.

\paragraph{Mitigating supply deficit}
We consider how different levels of flexibility affect the final-buyer supply deficit varying $\phi$ between 0 and 1.
We assign the same flexibility $\phi$ to all distributors.
In \cref{fig:deficit:phi_impact}, we show that flexibility considerably reduces the deficit of final buyers.
At $t$=40, the deficit decreases from about 6\% to 5\% as $\phi$ increases from 0 to 1.
This reduction means that about $500$k more Oxycodone doses are now reaching final buyers thanks to flexibility.
The largest reduction happens for some value $\phi^*$, corresponding to the $\phi$ value yielding the lowest supply deficit.
Importantly, we find that $\phi^*$ may be smaller than 1.

\paragraph{Acceptable Supply Deficit}

For essential goods, such as pharmaceuticals, a minimal supply level must be guaranteed.
We use the term \emph{acceptable supply deficit} (ASD) to refer to the maximum amount of goods that can be missing while still maintaining established standards.
In the case of Oxycodone, an acceptable supply deficit (ASD) would be the maximum deficit that does not compromise patient safety.
The concept of ASD is similar to that of service level agreements (SLAs), which set performance guarantees at the company level.
However, ASD differs from SLAs in that it is a systemic measure considering all final buyers.

Given an ASD, we define the \emph{resupply window} as the latest possible time $t$ at which resupply must happen before the deficit exceeds the ASD.
We find that the resupply window can be considerably extended thanks to flexibility.
In \cref{fig:deficit:resupplytime}, we show the maximum extension of the resupply window with flexibility for a given ASD. For small ASDs, the gain from flexibility is minimal. However, for larger values, the resupply window can be substantially extended.
For example, if a supply deficit of 5\% is acceptable, the resupply must happen within 20 days without flexibility. With enough flexibility, the resupply window can be extended by up to 38 days.

In \cref{fig:deficit:resupplygain}, we show the percentage gain that can be obtained for different levels of ASD.
We find that the resupply window can be extended by up to 80\%.
However, if the ASD is very low, e.g., 2\%, then this ASD will be breached quickly.
Thus, flexibility has no time to alleviate shortages.
If the ASD is very high, e.g., 10\%, when that supply deficit is reached, stocks will be depleted by regular demand.
Hence, we identify a range of ASD where flexibility is particularly effective.

\begin{figure}[t]
    \centering
    \begin{subfigure}{.4\textwidth}\centering
        \begin{subfigure}{\textwidth}\centering
            \includegraphics[width=\textwidth]{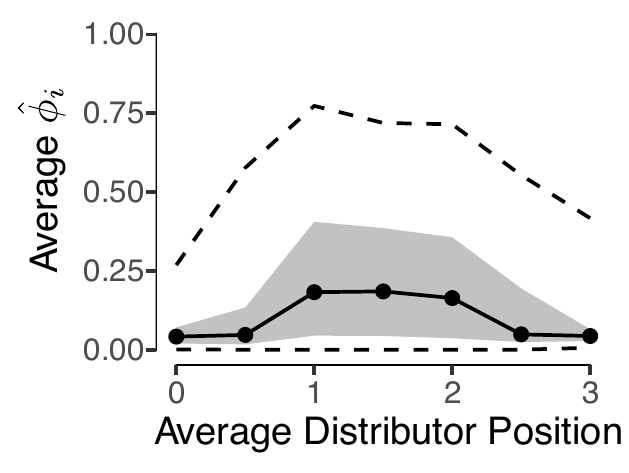}
            \caption{}\label{fig:phi_position}
        \end{subfigure}
        \begin{subfigure}{\textwidth}\centering
            \includegraphics[width=\textwidth]{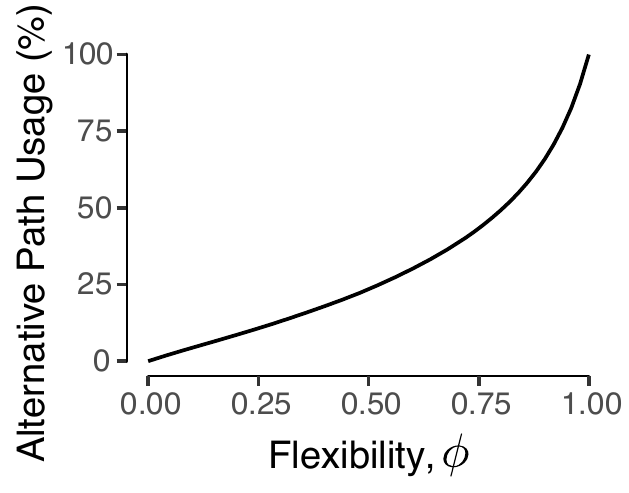}
            \caption{
            }\label{fig:hairball:usage}
        \end{subfigure}
    \end{subfigure}\hfill
    \begin{subfigure}{.6\textwidth}\centering
        \includegraphics[width=0.95\textwidth]{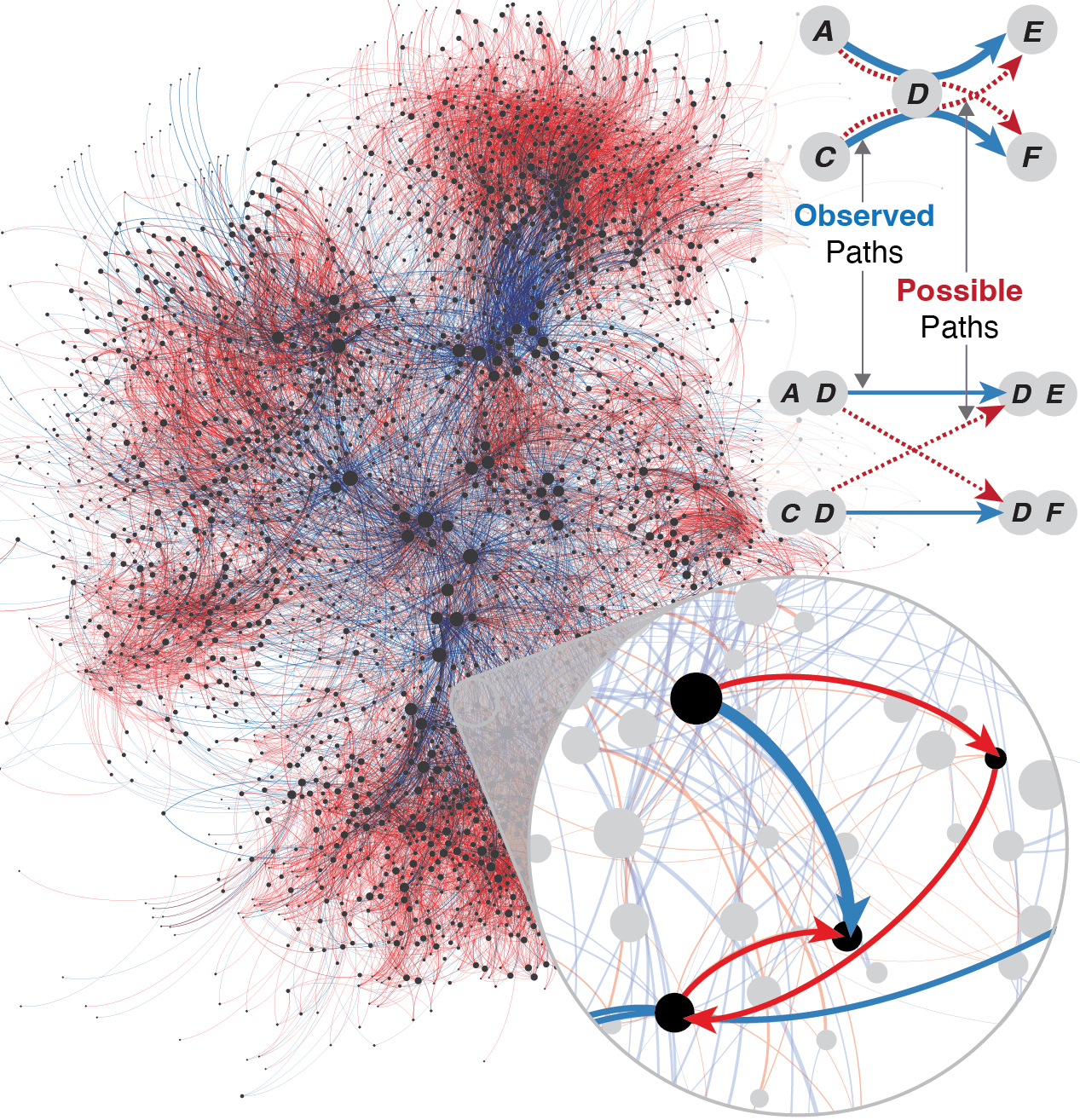}
        \caption{}\label{fig:hairball:ball}
    \end{subfigure}
    \caption{
        \textbf{(a)} Year-to-year flexibility $\hat{\phi}_i$ of distributor $i$ in the Oxycodone distribution system from 2006 to 2014, shown as a function of the distributor's position on the distribution paths. The black lines represent the average flexibility, the shaded area shows the middle 50\% of the data, and the dashed line shows 95\%.
        \textbf{(b)} Proportion of goods shipped via alternative distribution paths 180 days after the stop in production, for different levels of flexibility $\phi$,
        \textbf{(c)} The distribution system for Oxycodone in 2012 represented as a second-order network.
        In this representation, a path of length two, such as $A \to D \to E$, is depicted by an edge between the two ``meta-nodes'' $(A, D) \to (D, E)$ (as shown in the inset on the left).
        Blue edges indicate distribution paths that were observed, while the red edges represent alternative distribution paths that could exist.
        Increasing the parameter $\phi$ in this representation increases the probability that these red alternative paths become available for the distribution, in addition to the observed blue distribution paths.
        The zoom-in feature highlights that creating alternative distribution paths (red) allows previously disconnected nodes to connect.
        However, it is important to note that these alternative paths can be less direct than the observed blue paths, requiring products to follow longer routes.
    }
\end{figure}
\paragraph{Empirical Evidence for Flexibility}
Flexibility can mitigate supply deficits.
Now, we provide evidence that distributors can indeed adapt their upstream preferences and thus increase their flexibility.
We look at how empirical distribution systems evolve over time and assess the \emph{year-to-year} flexibility $\widehat{\phi}_{i}(y)$ of each distributor.
$\widehat{\phi}_{i}(y)$ captures how much distributor $i$ relaxes its upstream preferences from year $y-1$ to year $y$.
Precisely, we take a maximum likelihood approach (MLE) to infer $\widehat{\phi}_{i}(y)$ given the upstream preference in year $y-1$ and the observed distribution paths in year $y$.
See \cref{sec:methods:phihat} for details.

We find that, in every year, some degree of flexibility is present.
While on average distributors' flexibility is low, large flexibility values are sporadically observed.
To understand which distributors are more flexible, we compute the average position a distributor has on their distribution paths.
For example, distributor $D$ in \cref{fig:example} has position 2 in both the green and blue distribution path.
In \cref{fig:phi_position}, we see how the average $\widehat\phi_i$ changes with positions.
Distributors appearing at the beginning of paths have low flexibility, as do distributors at the end of paths, i.e., close to final buyers.
Instead, distributors occupying middle positions are more flexible, with an average $\widehat\phi_i(t)$ as high as 0.25.
In fact, 95\% of distributors occupying intermediate positions within the distribution system have a flexibility as high as 0.75.
This suggests that (i) distributors are able to adapt their preferences and (ii) maximum flexibility depends on their position.

\section*{Balancing deficit reduction and the cost of flexibility}
\paragraph{Flexibility introduces alternative distribution paths}
We compute the proportion of goods distributed through alternative paths as flexibility increases.
In \cref{fig:hairball:usage}, we see that the usage of alternative paths grows monotonously with $\phi$.
In other words, the more flexible, the more likely are distributors to use alternative distribution paths.
This allows final buyers to receive goods from multiple sources.

To understand where these alternative distribution paths are introduced, we visualise the distribution system in \cref{fig:hairball:ball}.
Blue edges show empirical distribution paths, while red edges represent the alternative paths available with full flexibility, i.e., $\phi=1$.
The zoom-in feature in \cref{fig:hairball:ball} shows that adding alternative paths (red) allows distribution between previously disconnected distributors.
Importantly, from \cref{fig:hairball:ball} we learn that the bulk of alternative distribution paths made available with flexibility is located towards the periphery of the distribution system.

\paragraph{The price of flexibility}
\begin{figure}[t]\centering
    \begin{subfigure}{.5\textwidth}\centering
        \includegraphics[width=\textwidth]{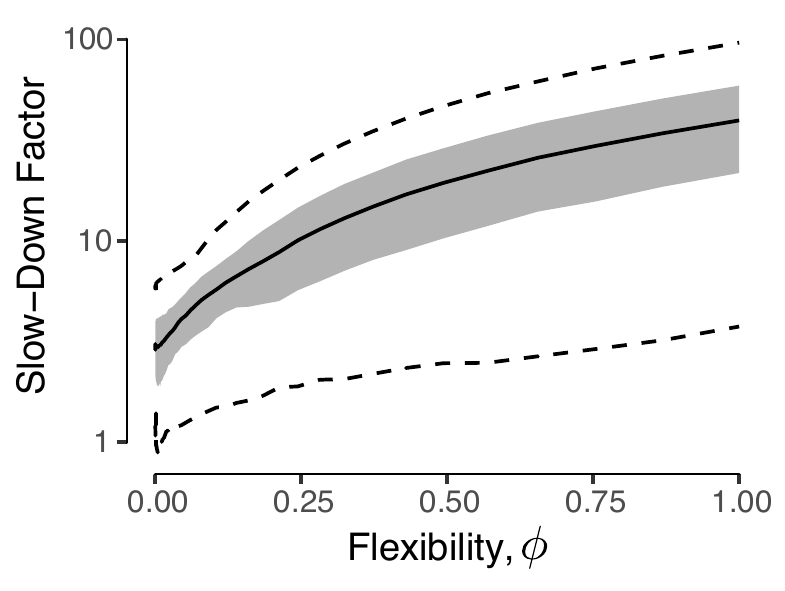}
        \caption{}\label{fig:fig2:slowdown}
    \end{subfigure}\hfill
    \begin{subfigure}{.5\textwidth}\centering
        \includegraphics[width=\textwidth]{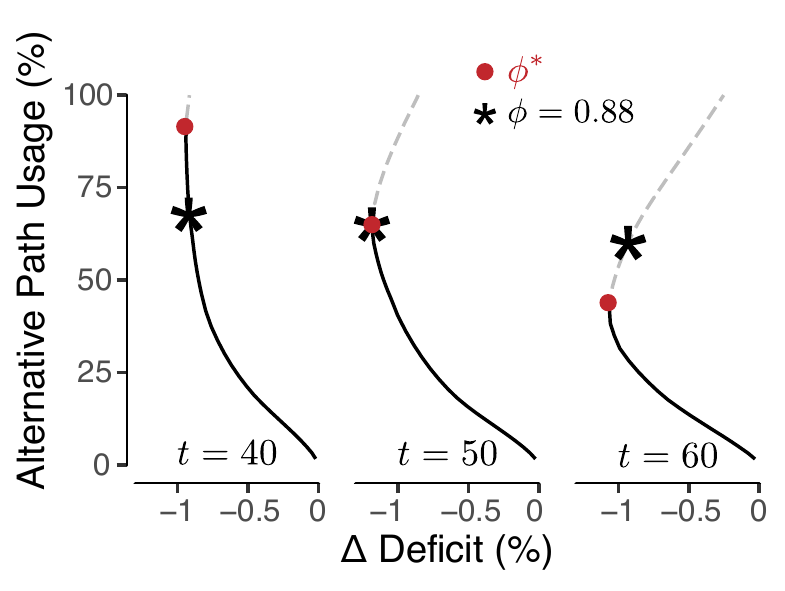}
        \caption{}\label{fig:fig3:efficient}
    \end{subfigure}
    \caption{
        \textbf{(a)} Slow-down factor for the Oxycodone distribution system, averaged from 2007 to 2014, as a function of $\phi$.
        The solid line indicates the average; the dashed lines show bootstrapped 95\% CI, and the shaded area indicates the 50\% CI.
        All bootstrapped statistics are estimated over 10 000 samples.
        \textbf{(b)} Deficit Reduction ($\Delta$) versus alternative path usage, plotted at 40, 50, and 60 days after production stop.
        $\phi$ increases from bottom to top along the curves.
        Solid lines show efficient values of $\phi$, while dashed lines show inefficient ones separated at $\phi^*(t)$.
        The point corresponding to $\phi^*(50)$ is shown in red, which is efficient at $t=40$, has the highest deficit reduction at $t=50$, and becomes inefficient at $t=60$.
    }\label{fig:fig2}
\end{figure}

The increase in path usage may be costly.
Costs may rise, for example, because of handling, labour, and increased complexity.
Using alternative paths may also slow down distribution because they can be less direct, as shown in the zoom-in feature in \cref{fig:hairball:ball}.
Therefore, alternative distribution paths may delay distribution to final buyers.

To proxy such a distribution delay, we compute the slowdown factor introduced by~\citet{scholtes2014causality}.
This factor indicates the proportion of additional distribution steps required for to reach final buyers.
By modeling the distribution of goods as a diffusion process, we are able to estimate how the  average distribution time scales with flexibility.
Details are provided in \cref{sec:methods:slowdown}.

In \cref{fig:fig2:slowdown}, we observe that increasing $\phi$ slows down the distribution system monotonously.
Thus, as flexibility increases, goods pass through more distributors, potentially raising handling costs.
This suggests the presence of a tradeoff between flexibility and the usage of alternative paths.

\paragraph{Decreasing returns to flexibility}
In \cref{fig:fig3:efficient}, we see the tradeoff between deficit reduction and alternative path usage.
The plot highlights the existence of an \emph{inefficient set} located on the upper side of the curves (dashed line).
For points in this set, a given deficit reduction is achievable at lower flexibility as well.
In other words, the same deficit is attainable using fewer distribution paths, and thus with lower costs.
The set of points where this happens is the efficient set.
Moreover, in \cref{fig:fig3:efficient} we show that an efficient point at $t=40$ becomes inefficient at $t=60$.
This occurs because the value of $\phi^*(t)$, separating the efficient from the inefficient set, decreases with time, which is also visible in \cref{fig:deficit}.

%% file: discussion.tex
Natural disasters, geopolitical tensions, and public health crises can severely disrupt supply chains, leading to shortages.
Our work demonstrates that flexibility, i.e., the ability to substitute goods through existing distribution paths, mitigates shortages and, crucially, extends the time before a critical deficit is reached.
Specifically, to quantify the distribution system's ability to mitigate supply deficits we developed a new analysis tool for  distribution systems using stochastic chains with memory \citep{rissanen1983,buhlmann1999}.

Strengthening supply chain resilience, i.e. the ability to withstand and recover from shocks~\citep{Hollnagel2007,Sterbenz2010}, was declared a top national security by US President Obama in 2012~\cite{obamabarack2012-nationalstrategy}.
To reconstitute the flow of commerce after disruption requires proactive and reactive measures.
Proactive measures strengthen the supply chain's ability to withstand shocks by taking preventive action \emph{before} disruptions occur~\cite{casiraghi2020_interventions,intervention_scenarios}, aiming to avert shortages altogether.
Examples of proactive measures include mandating higher safety stocks, investing in just-in-case capacity, and pursuing diversification~\cite{tucker2020-drugshortage}.
In contrast, reactive measures prioritize swift responses \emph{after} a shortage emerges, allowing the system to adapt and mitigate the effects of the disruption~\cite{fragile_yet_resilient,resilience_i_ching,resilience_mega_paper}.
While proactive measures may require significant upfront investments and, crucially, time, reactive measures are immediately available.

Flexibility, a reactive measure, leverages existing resources such as infrastructure, business relations, and goods, making them immediately available without creating new connections.
However, flexibility is costly due to increased handling time and distribution complexity.
Consequently, there is a tradeoff between its benefits and costs.
To manage this tradeoff and foster flexibility, regulators and policymakers must continuously monitor distribution paths, to gain insights into how flexibility can extend the time until a critical deficit is reached.

Our work provides the necessary tools to evaluate flexibility and stress-test the system continuously.
Our analysis has highlighted that
the most effective flexibility level changes with time and
the impact of flexibility is highest during the initial phase of a shortage.
This becomes important when devising policies to foster flexibility.
Our approach is applicable to a broad range of products, not just pharmaceuticals, and is well-suited for substitutable products with partially overlapping distribution systems, e.g., grain, gas and oil.
By carefully balancing policies that foster flexibility and costs, supply chains can become more resilient, enabling them to adapt to disruptions.

%% file: methods.tex
\subsection{The ARCOS dataset}
We study the opioid shipments dataset ARCOS~\cite{ARCOS}, which serves as our model distribution system.
The US Drug Enforcement Agency (DEA) maintains this dataset and tracks the chain of custody of every shipment of a controlled substance from manufacturing to the dispenser.
ARCOS, which stands for Automated Reports and Consolidated Ordering System, is a data collection system where manufacturers and distributors report controlled substance transactions to the DEA.
Transactions recorded in ARCOS include information such as the sending and receiving entities, the quantity and good shipped, and the date.
Drugs are identified by their national drug code, which allows us to distinguish the labeller, good, and packaging forms, such as 12ml vials or 120 pill boxes.
It is worth noting that the last entities to be tracked by ARCOS are pharmacies, hospitals, and practitioners, not the patients, and they will be referred to as ``final buyers'' hereafter.

\paragraph*{Substitutability and price elasticity}
The extent to which substitution alleviates shortages depends on the \emph{substitutability} of the products.
The FDA defines drugs to be ``pharmacologically equivalent'' if they contain (1) the same active ingredient, (2) have the same dosage form, and (3) are identical in strength and concentration.
In this work, we follow this definition but relax the ``identical in strength'' and ``dosage form'' requirements.

Although price differences can affect product substitutability, we have chosen not to model them in this study.
This decision is based on research by \citet{yeung2018price}, which found that medically necessary drugs, such as painkillers, are not significantly affected by short-term price increases.
Furthermore, price increases do not stimulate supply, as noted by the US Food and Drug Administration in their report on drug shortages~\cite{usfda2019-drugshortages}.
This is primarily due to the low price elasticity of prescription drugs, which is caused by how necessary pharmaceuticals are reimbursed.
Insurers and federal programs, rather than patients, are usually responsible for paying for these drugs.

\paragraph*{From shipping transactions to distribution paths}
In this study, we reconstruct the distribution paths of opioid drugs by tracking all ARCOS transactions in the order they were recorded while monitoring distributor stock levels.
Specifically, we trace individual packages as they leave the manufacturing facility, pass through distributors, and arrive at final buyers, e.g., hospitals, pharmacies, or practitioners.

To do so, we assume that distribution systems follow a first-in-first-out (FIFO) stock management policy, where the first packages arriving are also the first ones to leave.
This policy minimizes the impact on the product's shelf-life, which is crucial for perishable products such as medicine.
In fact, the World Health Organization recommends in their "Good Distribution Practices"~\cite{who2020good} that distributors follow a "first expiry/first-out" stock management policy.

Using the 500 million transactions in ARCOS for 2006-2014, we reconstruct 40 billion distribution paths of individual drug packages.
The set of reconstructed paths is denoted as $P := \{ p_1, p_2, \dotsc, p_S \}$, where each element in the set is a single path of a single drug package.
Each path is represented as a tuple $p_s = (M \to k \to j \cdots \to i)$.
Here, $M$ denotes a manufacturer and $k,j,i$ denotes distributors.
The path ends with the last distributor that ships to the final buyers.
In other words, $p_s$ denotes the sequence of distributors traversed by a given package on its journey from manufacturer to final buyer.

\subsection{Higher-order Markov chains model for distribution systems}
\paragraph*{From distribution paths to upstream preferences}
The length of reconstructed paths varies between 1 and 4.
The majority of these paths, though, has a length of 2.
This means that, in most cases, the distribution process involves only one manufacturer and two subsequent distributors.
Given this observation, we choose to model upstream preferences up to 2 steps upstream.
Supporting this assumption, \citet{hypa} have shown that this approach is particularly suitable to capture important paths within a system.
Further, we validate this modelling choice performing the model selection tests proposed by~\citep{scholtes2017KDD,mogen}.
The tests show that modelling the distribution system accounting for 2 steps upstream is statistically optimal, given the available data.

Let's consider the length-2 distribution path $p_s = (k \to j \to i)$, where $k$ is a manufacturer, and $j$ and $i$ two distributors.
In our data, each length-2 path may appear as a full observation or as a subpath of a longer distribution path, i.e., $(\cdots \to k \to j \to i \to \cdots)$.
We denote with $\tilde{A}_{kji}$ the total number of occurrences of $p_s$ in the data, summing all its occurrences as standalone path with those as subpath of longer distribution paths.

To proceed further, we assume perfect market clearing within the system.
Under this assumption, supply equals demand.
This implies that the amount of shipments corresponds to the orders placed.
This means that $\tilde{A}_{kji} = A_{ijk}$, where $A_{ijk}$ indicates the amount of orders placed by $i$ to $k$, via the intermediary $j$.
To model the distribution system, we leverage higher-order Markov chains and construct the 2-step tensor, $T^\text{2-step}$.
Each entry of $T^\text{2-step}_{ijk}$ contains the probability that $i$ submits an order to $k$ via the intermediary $j$:
\begin{equation}
    T^\text{2-step}_{ijk} = \frac{A_{ijk} }{ \sum\limits_{ j'k'}A_{ij'k'} }
    \label{eq:T_non_markov}
\end{equation}
where the sum runs over all sub-paths $(j' \to k')$.
Formally, each element of $T^\text{2-step}$ represents the \emph{transition probability} of an order moving along a path $(i\to j\to k)$.
$T^\text{2-step}$, thus, captures \emph{all} upstream preferences up to 2 steps upstream.
Note that \cref{eq:T_non_markov} ensures the dependency between the (probability of) orders placed by $i$ towards $j$  and the (probability of) orders placed by $j$ towards $k$, namely $P(i \to j \to k)$.

\paragraph*{Relaxing upstream preferences}

Distributors may relax their upstream preferences and, in the most extreme case, accept goods independently of their origin.
To capture this tendency, we introduce a 1-step transition matrix, $S$.
A given element $S_{ij}$ captures the probability that $i$ places an order to $j$.
Formally, we write $S_{ij} = \sum_{k} T^\text{2-step}_{ijk}$ where the sum runs over all distributors $k$.
Using this 1-step transition matrix, we construct a new tensor, $T^\text{1-step}$, that captures preferences up to 1 step upstream while modelling paths of length 2:
\begin{equation}
    T^\text{1-step}_{ijk}  = \frac{S_{jk}}{\sum\limits_{k'} S_{jk'}} \cdot \Theta\left(\sum\limits_{k'} A_{ijk'}\right)
    \label{eq:T__markov}
\end{equation}
where $\Theta(x)$ equals 0 for $x\leq0$ and equals 1 otherwise.
It ensures that we only considers a distributor $j$ if there is at least one order placed by $i$ towards $j$.
Note that, except for the $\Theta$, the right-hand side has only two $(jk)$ while the left-hand side of~\cref{eq:T__markov} has three indices $(ijk)$.
This is not a mistake.
We are assuming that $i$ has fully relaxed its upstream preferences, aligning them to the intermediary $j$.
As a consequence, the proportion of orders that $i$ places toward $k$ does not depend on $i$ anymore, but it only depends on the proportion of orders that $j$ places toward $k$.

\paragraph*{Flexibility}
Upstream preferences are relaxed according to various level of the distributors' flexibility.
To model different levels of flexibility, we combine the $T^\text{1-step}_{ijk}$ and the $T^\text{2-step}_{ijk}$ as:
\begin{equation}\label{eq:mixture_phi}
    T(\phi_i)_{ijk} = (1-\phi_i) T^\text{2-step}_{ijk} + \phi_i T^\text{1-step}_{ijk}
\end{equation}
where $\phi_i$ is a parameter used to interpolate between the two limit cases: (i) a fully flexible case captured by $T^\text{1-step}$, (ii) and a zero flexible system captured by $T^\text{2-step}$.
Its value ranges from zero to one and indicates the percentage of goods received by distributors independently of their upstream preferences.
When flexibility equals zero, $T(\phi_i=0)$ reduces to the 2-step tensor, i..e, $T^\text{2-step}$.
When flexibility equals one, $T(\phi_i=0)$ reduces to the 1-step tensor, i.e., $T^\text{1-step}$.

We can visualize both $T^\text{1-step}$ and $T^\text{2-step}$ using a second order graphical representation (\cref{fig:hairball:ball}).
In this representation, paths of length 2 $i \to j \to k$ are represented by an edge between the two ``meta-nodes'' $(i, j) \to (j, k)$ (see inset on the bottom-right of the figure).
This figure shows observed paths of length two in blue, i.e., the positive entries in $T^\text{2-step}$.
Paths, red,  are possible but have not been observed and correspond to the positive entries in $T^\text{1-step}$.
Increasing $\phi$ in this representation corresponds to adding red edges (possible paths) to the observed paths (blue edges).

\subsection{Estimating the empirical flexibility}
\label{sec:methods:phihat}

We use a maximum likelihood approach to estimate the system's empirical flexibility at a given time horizon $h$.
Specifically, we estimate the upstream preferences for each distributor by computing the shipment transition tensor $B(b, \pmb\phi)$ over a period $[t-b, t]$, where $b$ is the period over which the preferences are estimated and $\pmb\phi$ is an n-dimensional vector whose entries $\phi_i$ correspond the flexibility of distributor $i$.

Specifically, we obtain $B(b, \pmb\phi)$ as the row normalized transpose of the order transition tensor $T(\pmb\phi, b)$ defined in \eqref{eq:mixture_phi}.
The rationale behind this is that we define expected shipments to be equal to expected orders assuming each distributor has placed orders for the observed volume.
Formally,
\begin{equation}\label{eq:shiptrans}
    B_{ijk}(b, \phi_k) := \frac{T_{kji}(b, \phi_k) \cdot v_k }{\sum\limits_{k'j'} T_{k'j'i}(b, \phi_{k'})\cdot v_{k'}}
\end{equation}
where $T_{ijk}(b, \phi_k)$ is defined in~\eqref{eq:mixture_phi} over the period $[t-b, t]$ and $v_k=\sum_{lm}A_{klm}$ is the total volume ordered by $k$.

We then construct from the shipments observed in the period $[t, t+h]$ the shipment tensor $\tilde{A}(h)$.
The entry $\tilde{A}_{ijk}(h)$ captures the number of shipments from $i$ to $k$ via $j$ in the period $[t, t+h]$.
Finally, we compute the likelihood of the observed shipments given the estimated transition tensor $B(b, \pmb\phi)$ parametrized by $\pmb\phi$ as:
\begin{equation}
    \mathcal{L}(\phi) = \prod\limits_{i,j,k} B_{ijk}(b, \phi_k)^{\tilde{A}_{ijk}(h)} \propto
    \log \mathcal{L}(\phi) = \sum\limits_{i,j,k} \tilde{A}_{ijk}(h) \log B_{ijk}(b, \phi_k)
\end{equation}
The most likely parameter to have generated the observed shipments corresponds to the flexibility vector $\pmb{\hat{\phi}}$ for which the likelihood is maximal.
\begin{equation}
    \pmb{\hat{\phi}} = \arg\max\limits_{\pmb\phi} \log \mathcal{L}(\pmb\phi)
\end{equation}
In \cref{fig:phi_position}, we estimate upstream preferences over a year ($b=$ 1 year) and then use the estimated transition matrix to predict the shipments over the next year ($h=$ 1 year).

\subsection{Slow-down factor}\label{sec:methods:slowdown}
To compare the distribution speed between flexible and strict upstream preferences, we employ the slow-down factor introduced by \citet{scholtes2014causality}.
Let $M$ denote a row-stochastic transition matrix describing a random walk on a network.
It has been shown that the time $t$ needed for the node visitation probability to converge to the stationary distribution starting from any initial condition scales with
\begin{equation}\label{eq:convergtime}
    t \approx \frac{1}{\log{\abs{\lambda_2\left[M\right]}}}\,,
\end{equation}
where $\lambda_2\left[M\right]$ is the second leading eigenvalue of $M$.

Consider now the $B(b,\pmb\phi)$ tensor defined in \cref{eq:shiptrans} and by setting the vector $v=\mathbb{1}$.
Its elements $B_{ijk}(b,\pmb\phi)$ are the probabilities of a shipment from $i$ to $k$ via $j$ as a function of flexibility $\pmb\phi$.
We can map the $B(b,\pmb\phi)$ $n\times n\times n$ tensor representing 2-steps transitions to an equivalent $n^2 \times n^2$ second-order transition matrix $\tilde B(b,\pmb\phi)$ as follows.
A second-order node $(i,j)$ denotes that $i$ ships to $j$ in the distribution system.
If the shipment from $i$ to $j$ does not exists, the second-order node $(i,j)$ does not exist~\cite{scholtes2014causality,scholtes2017KDD}.
In the other cases,
\begin{equation}
    \tilde B_{(i,j)(m,k)}(b,\pmb\phi) =
    \begin{cases}
        B_{ijk}(b,\pmb\phi) \text{ iff $m=j$,} \\
        0 \text{ otherwise.}
    \end{cases}
\end{equation}

Let $\Omega$ denote the set of \emph{final distributors}, i.e., of distributors that ship goods downstream to final buyers (patients, hospital, pharmacies).
By connecting each final distributor $\omega\in\Omega$ to an \emph{end-node} $\dagger$, we can model the fact that distribution paths end at final distributors (see \citep{mogen} for more details):
\begin{equation}
    \tilde B_{(\omega,\dagger)(\dagger)}(\phi) > 0 \text{ iff $\omega\in\Omega$,}
\end{equation}
where, with an abuse of notation, we denote with $(\dagger)$ the second-order representation of the end-node.
Finally, we set $\tilde B_{(\dagger)(\dagger)}(b,\pmb\phi)=1 \forall \pmb\phi$.
By doing so, we ensure that the Markov-chain defined by $\tilde B(b,\pmb\phi)$ is absorbing and has a unique stationary distribution $(0,\dots,0,1)$, where the last element corresponds to the end-node $(\dagger)$.
Thus, all random walks converge to $(\dagger)$.

Let $\lambda_2\left[\tilde B(b,\pmb\phi)\right]$ be the second leading eigenvalue of $\tilde B(b,\pmb\phi)$.
Then, we can define the slow-down factor $\sigma(\pmb\phi)$ as the additional number of steps it takes for the visitation probability to converge to its stationary distribution compared to the reference case $\pmb\phi=\mathbb{0}$.
Formally, from \cref{eq:convergtime}:
\begin{equation}
    \sigma(\phi) := \frac{\log\abs{\lambda_2\left[\tilde B(b,\mathbb{0})\right]}}{\log\abs{\lambda_2\left[\tilde B(b,\pmb\phi)\right]}}\,.
\end{equation}
The full derivation of this result is provided by \cite{scholtes2014causality}.

\subsection{Modeling distribution dynamics with upstream preferences}
\label{sec:methods:ARIO}

When a shock hits the distribution system, it can response to it with various levels of flexibility.
To model how distribution dynamics change depending on the level of flexibility considered, we extend the ARIO (Adaptive Regional Input Output) model introduced by~\citep{hallegatte2014modeling}.
In this extension, we propose to incorporate the distributors' upstream preferences.
According to the ARIO principles, distributors place orders to (i) meet demand and (ii) avoid empty inventories by keeping them at a constant target level, $s^T$, or safety buffer, i.e., :
\begin{equation}
    o_{(i|j)}(t) = d_{(i|j)} \left( t-1 \right) + \frac{1}{\tau} \left[ s_{(i|j)}^{ T} - s_{(i|j)}(t)\right]
    \label{eq:order-dynamic-i}
\end{equation}
In \cref{eq:order-dynamic-i} $o_{(i|j)}$ is the order placed by $i$ towards $j$ and $d_{(i|j)}$ is the demand $i$ faces on the goods received from $j$.
The demand $d_{(i|j)}$ takes into account two terms: orders received from (a) final buyers and (b) orders received from other distributors.
The term (a) is captured by the vector $c$, the term (b) is captured by the order matrix $O$.
Following~\citet{hallegatte2014modeling}, we model the two terms separately.

The parameter $\tau$ indicates how quickly distributor $i$ wants to restore its inventories.
To keep our model simple, we consider $\tau$ homogenous across distributors and constant over time.
In our study, we set it equal to one working week, i.e., $\tau=5$ days.

$s_{(i|j)}$  represents the sub-stock of $i$ used to store goods received from $j$.
Note that unlike the original version of the ARIO model, in the presented model distributors hold stocks divided into sub-stocks.
A substock $s_{(i|j)}$ represents the part of the stock used by $i$ to store goods coming from $j$.
In this way, we keep track of the stage before the goods enter the warehouse.
Sub-stocks are updated according to the total ship-out and the total ship-in:
\begin{equation}
    s_{(i|j)}(t) = s_{(i|j)}(t-1)  +  W_{(i|j)}^{ \mathrm{in}}(t-1)   - \left[ W_{(i|j)}^{  \mathrm{out}}(t-1) + \omega_{(i|j)} (t-1) \right]
    \label{eq:stock-dynamic_2order}
\end{equation}
The second term on the right-hand side indicates the total amount of goods $i$ received from $j$.
The third term, i.e., the one in parenthesis, indicates the amount of goods shipped by $i$ given that it has received such goods from $j$.
This total ship-out captures both the amount directed to final buyers, $\omega_{(i|j)}$, and the amount directed to other distributors, $W_{(i|j)}^{ \mathrm{out}}$.

Once stocks are updated, distributors places orders while respecting their upstream preferences captured by the tensor $T_{ijk}$ as:
\begin{equation}
    O_{ijk}(t)= o_{(i|j)}(t)  T_{ijk} \left( \phi\right)
    \label{eq:order_ijk}
\end{equation}
where $T(\phi)$ is defined in~\cref{eq:mixture_phi}.
Thus, in the case of zero flexibility, $\phi=0$, upstream preferences are kept fix.
In the case of medium flexibility, $\phi\neq 0$, upstream preferences are relaxed.

Finally, assuming that distributors want to meet demand as much as possible.
The quantity shipped by a given distributor $i$ is always determined as the maximum between the orders faced by $i$ and its stock level.

\subsection{Initializing the model with real-world data}

In a stress-test approach, we want to start with the closest representation of the real system, and simulate its deviation given a possible supply shock.
Building on this reasoning, we initialize the demand from final buyers and the stock levels of distributors with the empirical data.
First, we assume that distributors meet demand perfectly within the observation year, $y$.
Based on this assumption, we determine the constant daily demand faced by distributor $i$ as:
\begin{equation}
    c_{i} =  \frac{ \omega_{i}(y) }{365}
\end{equation}
where $\omega_{i}(y)$ indicates the amount $i$ shipped to final buyers in the year $y$.
Then, respecting the proportion of volumes observed, we obtain the demand faced by $i$ and conditioned to distributor $j$ as:
\begin{equation}
    c_{(i|j)} =   c_{i}  \frac{ W^{\mathrm{in}}_{(i|j)}(y)}{\sum\limits_{j'} W^{\mathrm{in}}_{(i|j')}(y) }
\end{equation}

Next, we determine the target stocks assuming that all distributors meet their planning within the observation year.
Under this assumption, the target stocks are obtained as the empirical buffer observed at the end of the year\footnote{Note that, in some cases, the ship-out is bigger than the ship-in.
    This suggests that: (i) their inventories were not empty at the beginning of the given year, or (ii) they did not plan a target (safety) stock.
    For these distributors, we set a minimum buffer equal to one.}, as:
\begin{equation}
    s^{T}_i = W_i^{\mathrm{in}}(y) - \left[ W_i^{\mathrm{out}}(y) -  \omega_{i}(y)  \right]
    \label{eq:stock-target-1order}
\end{equation}
where the first term on the right-hand side indicates the total ship-in of $i$ in the year $y$; whereas the second term indicates the total ship-out of $i$ in the year $y$.
Then, respecting the proportion of volumes observed, we compute the target sub-stock of $i$ conditioned to distributor $j$ as:
\begin{equation}
    s^{T}_{(i|j)} = s^{T}_i \frac{ W^{\mathrm{in}}_{(i|j)}(y)}{\sum\limits_{j'} W^{\mathrm{in}}_{(i|j')}(y) }
\end{equation}
All stocks are initialized to their target values at the beginning of the simulation.

\subsection{Simulating a supply shock}
We consider an external shock that reduce the total production by $\sigma$ percentage, i.e, :
\begin{equation}
    s_{i}(t=t^*) =   (1-\sigma) s_{i}(t-1) \quad \forall i\in \lbrace m_1, m_1, \dots, m_n \rbrace
\end{equation}
where $\sigma$ is the size of the shock and $s_{i}$ denotes the manufacturer's stock level (used to store its production), and $t^*$ is the time the shock hits the system.

\subsection{Measuring supply deficit}
To evaluate the effect of flexibility in mitigating shortages we measure the reduction of supply \emph{deficit} for final buyers.
Specifically, we define supply deficit, $\delta(t)$, at time $t$, as the percentage of the (cumulative) unfulfilled demand of final buyers, i.e.:
\begin{equation}
    \delta(t)  = \frac{ \sum\limits_{t'=0}^{t}  \sum\limits_{i}  \omega_{i} (t') -  c_{i} }{ t \times \sum\limits_{i} c_{i} }
    \label{eq:measure-demand-deficit}
\end{equation}
where $i$ runs over all distributors shipping to final buyers.
Our indicator is built assuming that goods ordered are shipped within the next working day.

\subsection{Measuring alternative path usage}
Flexibility  introduces alternative distribution paths that can be used to source substitutable goods and mitigate the shortage.
Therefore, flexibility brings changes to usual operations resulting from a difference in the usage of the distribution paths.
To quantify such changes, we consider the amount shipped in two scenario: when flexibility is zero and when it is different from zero.
The difference between those two quantities gives the difference in the amount of goods shipped between every distributor pair when upstream preferences are relaxed.
We normalize such absolute difference with the maximum possible difference, occurring for $\phi=1$, thus obtaining:
\begin{equation}
    \Gamma(t) = \frac{\sum\limits_{ij} \abs{ W_{(i|j)} (\phi, t) - W_{(i|j)}(\phi=0, t)}}{\sum\limits_{ij} \abs{ W_{(i|j)}(\phi=1, t) - W_{(i|j)}(\phi=0, t)}}
    \label{eq:measure-change-system}
\end{equation}

%% file: main.bbl
\begin{thebibliography}{33}
\expandafter\ifx\csname natexlab\endcsname\relax\def\natexlab#1{#1}\fi
\expandafter\ifx\csname url\endcsname\relax
  \def\url#1{\texttt{#1}}\fi
\expandafter\ifx\csname urlprefix\endcsname\relax\def\urlprefix{URL }\fi
\expandafter\ifx\csname selectlanguage\endcsname\relax
  \def\selectlanguage#1{\relax}\fi

\bibitem[{Bier \emph{et~al.}(2020)Bier, Lange and
  Glock}]{bier2020-methodsmitigating}
Bier, T.; Lange, A.; Glock, C.~H. (2020).
\newblock Methods for Mitigating Disruptions in Complex Supply Chain
  Structures: A Systematic Literature Review.
\newblock \emph{International Journal of Production Research} \textbf{58(6)},
  1835--1856.

\bibitem[{Buhlmann and Wyner(1999)}]{buhlmann1999}
Buhlmann, P.; Wyner, A.~J. (1999).
\newblock Variable Length Markov Chains.
\newblock \emph{The Annals of Statistics} \textbf{27(2)}, 480--513.

\bibitem[{Burkholz and Schweitzer(2019)}]{burkholz2019-internationalcrop}
Burkholz, R.; Schweitzer, F. (2019).
\newblock International Crop Trade Networks: {{The}} Impact of Shocks and
  Cascades.
\newblock \emph{arXiv:1901.05872 [nlin, physics:physics, q-fin]} .

\bibitem[{Casiraghi and Schweitzer(2020)}]{casiraghi2020_interventions}
Casiraghi, G.; Schweitzer, F. (2020).
\newblock Improving the Robustness of Online Social Networks: A Simulation
  Approach of Network Interventions.
\newblock \emph{Frontiers in Robotics and AI} \textbf{7}.

\bibitem[{Dolgui \emph{et~al.}(2018)Dolgui, Ivanov and
  Sokolov}]{dolgui2018-rippleeffect}
Dolgui, A.; Ivanov, D.; Sokolov, B. (2018).
\newblock Ripple Effect in the Supply Chain: An Analysis and Recent Literature.
\newblock \emph{International Journal of Production Research} \textbf{56(1-2)},
  414--430.

\bibitem[{Gote \emph{et~al.}(2020)Gote, Casiraghi, Schweitzer and
  Scholtes}]{mogen}
Gote, C.; Casiraghi, G.; Schweitzer, F.; Scholtes, I. (2020).
\newblock Predicting Sequences of Traversed Nodes in Graphs using Network
  Models with Multiple Higher Orders.
\newblock \emph{arXiv:2007.06662} .

\bibitem[{Hallegatte(2014)}]{hallegatte2014modeling}
Hallegatte, S. (2014).
\newblock Modeling the role of inventories and heterogeneity in the assessment
  of the economic costs of natural disasters.
\newblock \emph{Risk analysis} \textbf{34(1)}, 152--167.

\bibitem[{Hollingsworth and Herndon(2018)}]{hollingsworth2018-parenteralopioid}
Hollingsworth, H.; Herndon, CPE, C. (2018).
\newblock The Parenteral Opioid Shortage: {{Causes}} and Solutions.
\newblock \emph{Journal of Opioid Management} \textbf{14(2)}, 81.

\bibitem[{Hollnagel \emph{et~al.}(2007)Hollnagel, Woods and
  Leveson}]{Hollnagel2007}
Hollnagel, E.; Woods, D.~D.; Leveson, N. (2007).
\newblock \emph{{Resilience engineering: Concepts and precepts}}.
\newblock Ashgate Publishing.

\bibitem[{Ivanov \emph{et~al.}(2014)Ivanov, Sokolov and
  Dolgui}]{ivanov2014-rippleeffect}
Ivanov, D.; Sokolov, B.; Dolgui, A. (2014).
\newblock The {{Ripple}} Effect in Supply Chains: Trade-off
  `Efficiency-Flexibility-Resilience' in Disruption Management.
\newblock \emph{International Journal of Production Research} \textbf{52(7)},
  2154--2172.

\bibitem[{{Jongh, Thyra de} \emph{et~al.}(2021){Jongh, Thyra de}, {Becker,
  Dominik}, {Boulestreau, Mathieu}, {Dav\'e, Anoushka}, {Dijkstal, Felix},
  {King, Robert}, {Petrosova, Liana}, {Varnai, Peter}, {Vis, Christiaan},
  {Spit, Wim}, {Moulac, Maxime} and {Pelsya,
  Florent}}]{jonghthyrade2021-futureproofingpharmaceutical}
{Jongh, Thyra de}; {Becker, Dominik}; {Boulestreau, Mathieu}; {Dav\'e,
  Anoushka}; {Dijkstal, Felix}; {King, Robert}; {Petrosova, Liana}; {Varnai,
  Peter}; {Vis, Christiaan}; {Spit, Wim}; {Moulac, Maxime}; {Pelsya, Florent}
  (2021).
\newblock \emph{Future-Proofing Pharmaceutical Legislation: Study on Medicine
  Shortages}.
\newblock \emph{Tech. rep.}, {European Commission. Directorate General for
  Health and Food Safety.}, {LU}.

\bibitem[{LaRock \emph{et~al.}()LaRock, Nanumyan, Scholtes, Casiraghi,
  Eliassi-Rad and Schweitzer}]{hypa}
LaRock, T.; Nanumyan, V.; Scholtes, I.; Casiraghi, G.; Eliassi-Rad, T.;
  Schweitzer, F. ().
\newblock HYPA: Efficient Detection of Path Anomalies in Time Series Data on
  Networks.
\newblock In: \emph{Proceedings of the 2020 SIAM International Conference on
  Data Mining (SDM)}. Society for Industrial and Applied Mathematics, pp.
  460--468.

\bibitem[{Miller \emph{et~al.}(2021)Miller, Young, Dobrow and
  Shojania}]{miller2021-vulnerabilitymedical}
Miller, F.~A.; Young, S.~B.; Dobrow, M.; Shojania, K.~G. (2021).
\newblock Vulnerability of the Medical Product Supply Chain: The Wake-up Call
  of {{COVID-19}}.
\newblock \emph{BMJ Quality \& Safety} \textbf{30(4)}, 331--335.

\bibitem[{Obama(2012)}]{obamabarack2012-nationalstrategy}
Obama, B. (2012).
\newblock National Strategy for Global Supply Chain Security.
\newblock https://www.dhs.gov/national-strategy-global-supply-chain-security.

\bibitem[{Rissanen(1983)}]{rissanen1983}
Rissanen, J. (1983).
\newblock A universal data compression system.
\newblock \emph{IEEE Transactions on Information Theory} \textbf{29(5)},
  656--664.

\bibitem[{Scholtes(2017)}]{scholtes2017KDD}
Scholtes, I. (2017).
\newblock When is a Network a Network? Multi-Order Graphical Model Selection in
  Pathways and Temporal Networks.
\newblock In: \emph{Proceedings of the 23rd ACM SIGKDD International Conference
  on Knowledge Discovery and Data Mining}. KDD '17, New York, NY, USA:
  Association for Computing Machinery, p. 1037–1046.

\bibitem[{Scholtes \emph{et~al.}(2014)Scholtes, Wider, Pfitzner, Garas, Tessone
  and Schweitzer}]{scholtes2014causality}
Scholtes, I.; Wider, N.; Pfitzner, R.; Garas, A.; Tessone, C.~J.; Schweitzer,
  F. (2014).
\newblock Causality-driven slow-down and speed-up of diffusion in non-Markovian
  temporal networks.
\newblock \emph{Nature communications} \textbf{5(1)}, 1--9.

\bibitem[{Schweitzer(2022)}]{resilience_i_ching}
Schweitzer, F. (2022).
\newblock Group relations, resilience and the I Ching.
\newblock \emph{Physica A: Statistical Mechanics and its Applications}
  \textbf{603}, 127630.

\bibitem[{Schweitzer \emph{et~al.}(2022)Schweitzer, Andres, Casiraghi, Gote,
  Roller, Scholtes, Vaccario and Zingg}]{resilience_mega_paper}
Schweitzer, F.; Andres, G.; Casiraghi, G.; Gote, C.; Roller, R.; Scholtes, I.;
  Vaccario, G.; Zingg, C. (2022).
\newblock Modeling Social Resilience: Questions, Answers, Open Problems.
\newblock \emph{Advances in Complex Systems} \textbf{25(08)}, 2250014.

\bibitem[{Schweitzer \emph{et~al.}(2021)Schweitzer, Casiraghi, Tomasello and
  Garcia}]{fragile_yet_resilient}
Schweitzer, F.; Casiraghi, G.; Tomasello, M.~V.; Garcia, D. (2021).
\newblock Fragile, Yet Resilient: Adaptive Decline in a Collaboration Network
  of Firms.
\newblock \emph{Frontiers in Applied Mathematics and Statistics} \textbf{7}, 6.

\bibitem[{Schweitzer \emph{et~al.}(2020)Schweitzer, Zhang and
  Casiraghi}]{intervention_scenarios}
Schweitzer, F.; Zhang, Y.; Casiraghi, G. (2020).
\newblock Intervention scenarios to enhance knowledge transfer in a network of
  firms.
\newblock \emph{Frontiers in Physics} \textbf{8}, 382.

\bibitem[{SLCG Economic~Consulting(2019)}]{ARCOS}
SLCG Economic~Consulting, S. (2019).
\newblock Opioid {{Data}}.
\newblock https://www.slcg.com/opioid-data/.

\bibitem[{Sterbenz \emph{et~al.}(2010)Sterbenz, Hutchison, {\c{C}}etinkaya,
  Jabbar, Rohrer, Sch{\"{o}}ller and Smith}]{Sterbenz2010}
Sterbenz, J. P.~G.; Hutchison, D.; {\c{C}}etinkaya, E.~K.; Jabbar, A.; Rohrer,
  J.~P.; Sch{\"{o}}ller, M.; Smith, P. (2010).
\newblock {Resilience and survivability in communication networks: Strategies,
  principles, and survey of disciplines}.
\newblock \emph{Computer Networks} \textbf{54(8)}, 1245--1265.

\bibitem[{Tang and Tomlin(2008)}]{tang2008-powerflexibility}
Tang, C.; Tomlin, B. (2008).
\newblock The Power of Flexibility for Mitigating Supply Chain Risks.
\newblock \emph{International Journal of Production Economics} \textbf{116(1)},
  12--27.

\bibitem[{Trent and Monczka(2005)}]{trent2005achieving}
Trent, R.~J.; Monczka, R.~M. (2005).
\newblock Achieving excellence in global sourcing.
\newblock \emph{MIT Sloan Management Review} .

\bibitem[{Tucker \emph{et~al.}(2020)Tucker, Cao, Fox and
  Sweet}]{tucker2020-drugshortage}
Tucker, E.~L.; Cao, Y.; Fox, E.~R.; Sweet, B.~V. (2020).
\newblock The {{Drug Shortage Era}}: {{A Scoping Review}} of the {{Literature}}
  2001\textendash 2019.
\newblock \emph{Clinical Pharmacology \& Therapeutics} \textbf{108(6)},
  1150--1155.

\bibitem[{Urahn \emph{et~al.}(2017)Urahn, Coukell, Jungman, Snyder, Bournas and
  Kourti}]{urahn2017-drugshortages}
Urahn, S.~K.; Coukell, A.; Jungman, E.; Snyder, E.; Bournas, J.~E.; Kourti, T.
  (2017).
\newblock \emph{Drug {{Shortages}}}.
\newblock \emph{Tech. rep.}, {The Pew Charitable Trusts and International
  Society for Pharmaceutical Engineering}.

\bibitem[{USFDA(2019)}]{usfda2019-drugshortages}
USFDA (2019).
\newblock \emph{Drug {{Shortages}}: {{Root Cause}} and {{Potential
  Solutions}}}.
\newblock \emph{Tech. rep.}, {U.S. Food and Drug Administration}.

\bibitem[{Wang and Downing(2019)}]{wang2019-crisisepidemic}
Wang, B.; Downing, N.~L. (2019).
\newblock A Crisis within an Epidemic: Critical Opioid Shortage in {{US}}
  Hospitals.
\newblock \emph{Postgraduate Medical Journal} \textbf{95(1127)}, 515--516.

\bibitem[{Whitehouse(2021)}]{thewhitehouse2021-nationalstrategy}
Whitehouse, T. (2021).
\newblock \emph{National {{Strategy}} for a {{Resilient Public Health Supply
  Chain}}}.
\newblock \emph{Tech. rep.}, {The Whitehouse}.

\bibitem[{WHO(2020)}]{who2020good}
WHO (2020).
\newblock \emph{{{TRS}} 1025 - {{Annex}} 7: {{Good}} Storage and Distribution
  Practices for Medical Products}.
\newblock \emph{Tech. rep.}, {World Health Organisation}.

\bibitem[{Woodcock and Wosinska(2013)}]{woodcock2013-economictechnological}
Woodcock, J.; Wosinska, M. (2013).
\newblock Economic and Technological Drivers of Generic Sterile Injectable Drug
  Shortages.
\newblock \emph{Clinical Pharmacology and Therapeutics} \textbf{93(2)},
  170--176.

\bibitem[{Yeung \emph{et~al.}(2018)Yeung, Basu, Hansen and
  Sullivan}]{yeung2018price}
Yeung, K.; Basu, A.; Hansen, R.~N.; Sullivan, S.~D. (2018).
\newblock Price elasticities of pharmaceuticals in a value based-formulary
  setting.
\newblock \emph{Health economics} \textbf{27(11)}, 1788--1804.

\end{thebibliography}
